\documentclass[12pt]{article}

\usepackage{amsfonts}
\usepackage{epsfig,amssymb,euscript}
\usepackage{amsmath,amscd}

\numberwithin{equation}{section}
\begin{document}

\begin{titlepage}
\begin{center}

\vspace*{-1.0cm}

%\hfill UB-ECM-PF-06-43 \\
\hfill DMUS-MP-18-01
\\
\vspace{2.0cm}

\renewcommand{\thefootnote}{\fnsymbol{footnote}}
{\Large{\bf Solutions of $D=4$ Gauged Pseudo-Supergravity}}
\vskip1cm
\vskip 1.3cm
J. B. Gutowski$^1$  and W. A. Sabra$^2$
\vskip 1cm
{\small{\it
$^1$Department of Mathematics,
University of Surrey \\
Guildford, GU2 7XH, UK.\\}}
\vskip .6cm {\small{\it
$^2$ Centre for Advanced Mathematical Sciences and Physics Department\\
American University of Beirut\\ Lebanon  \\}}
\end{center}
\bigskip
\begin{center}
{\bf Abstract}
\end{center}
 The techniques of spinorial geometry are used to classify solutions admitting Killing spinors in the theory of  minimal 
anti-de Sitter $N=2$, $D=4$ supergravity, where the gauge kinetic term comes with the opposite sign. 
There are four classes of solutions. One class is described by metrics corresponding to gravitational waves propagating on
$AdS_2\times H^2$. The second class of solution is a new solution corresponding to a special limiting 
case of the Killing spinor. The third class of solution corresponds to fibrations over a Lorentzian three 
dimensional manifold which has a Lorentzian Gauduchon-Tod structure. The fourth class of solution is a cosmological extension of a Majumdar-Papapetrou type solution,  
described by a function satisfying the wave equation on $\mathbb{R}^{2,1}$ .
\end{titlepage}

\section{Introduction}

In recent years, the study of supersymmetric background solutions of
supergravity theories had a great impact on the development of string and
M-theory. Moreover anti-de Sitter spaces became relevant to the study of
quantum field theories via the ADS/CFT\ correspondence. Supersymmetric
solutions are those bosonic gravitational backgrounds admitting a number of
Killing spinors $\epsilon $, satisfying $D_{\mu }\epsilon =0$, where $D_{\mu
}$ denotes the supercovariant derivative. Additional Killing spinor
equations arise in theories with scalar fields resulting from the vanishing
of supersymmetry transformation of additional fermionic fields other than
the gravitini.

In our present work we will consider minimal
supergravity theories in four dimensions. The first systematic
classification of solutions in minimal four-dimensional $N=2$ supergravity
without a cosmological constant was performed in \cite{Tod}. The
half-supersymmetric solutions found by Tod consist of two classes of
solutions. The first class has a time-like Killing vector and corresponds to 
an Israel-Wilson-Perjes metric, and the second class has a light-like Killing vector
with a plane-wave metric. Some analysis of the purely gravitational
backgrounds, i.e., without a $U(1)$ gauge field or a cosmological constant
can be found in \cite{Bryant, jose}. In \cite{klemm}, 1/4-supersymmetric and
1/2-supersymmetric backgrounds using the method of spinorial geometry were
analysed for gauged four-dimensional $N=2$ gauged supergravity. The bosonic
part of $N=2,D=4$ gauged supergravity is basically Einstein-Maxwell theory
with a negative cosmological constant. Preons of this theory were also
analysed in \cite{preonone}, where local analysis excluded simply connected
solutions preserving exactly 3/4 of the supersymmetry. Later in 
\cite{preontwo}, it was shown that 3/4-BPS preons in four-dimensional gauged
supergravities can be obtained as smooth quotients of the $AdS_{4}$
maximally supersymmetric backgrounds. This indicates that space-time
topology can in some cases resurrect solutions ruled out by the local
analysis of the Killing spinor equations.

 Einstein-Maxwell theory with a
positive cosmological constant cannot be embedded in a supergravity theory,
as supersymmetry restricts the cosmological constant to be negative.
However, with a positive cosmological constant, one can nevertheless construct a Killing
spinor equation by analytic continuation. The fake supersymmetry thus
obtained can be used as a solution generating technique \cite{ds}. It should
be mentioned that de Sitter supergravities were obtained via the reduction
of M*-theory and type IIB* theory \cite{dsreduction}. Recently, solutions
with space-like Killing vectors of $N=2,D=4$ supergravity, where gauge
field kinetic terms come with the opposite sign in the action, were
obtained in \cite{fake}. It must be noted that supergravity theories with
various space-time signatures and either sign of gauge field kinetic terms,
were recently explored in terms of reduction of * theories \cite{ch} on
Calabi-Yau manifolds \cite{sig}.

In our present work we are interested in the classification of solutions
admitting a Killing spinor in  minimal gauged $N=2,$ $D=4$
pseudo-supergravity. This theory can be obtained by taking standard $%
N=2$, $D=4$ minimal gauged supergravity, as set out in e.g. \cite{klemm},
and reversing the sign of the gauge kinetic term in the action. The
techniques of spinorial geometry developed for the analysis of supergravity
Killing spinor equations \cite{Gillard:2004xq} will be used to 
analyse the solutions.

The plan of the paper is as follows. In
section two, we summarise the basic equations of the theory we will be
studying and give a brief presentation of the orbits of the Killing spinors. In
sections three, four and five, we give a detailed analysis of the Killing
spinor equations for the three canonical forms of the Dirac spinors as well
as an explicit construction of their corresponding solutions. In all these
sections, we make use of the linear system presented in the Appendix. We
conclude in section six.

\section{$N=2$, $D=4$ Pseudo-Supergravity}

In this section, we present a summary of gauged $N=2$, $D=4$
pseudo-supergravity. As already mentioned, this theory is obtained from the
standard $N=2$, $D=4$ minimal gauged supergravity by rescaling the gauge
field as 
\begin{equation}
A\rightarrow -iA\text{ }.
\end{equation}
The bosonic action of the theory is then given by  
\begin{equation}
S=\int \ \mathrm{d}^{4}x\sqrt{-g}\ \left( R+F_{\mu \nu }F^{\mu \nu }+{\frac{6
}{\ell ^{2}}}\right)   \label{ac}
\end{equation}
where $\ell $ is a non-zero real constant. The signature of the metric is $%
(-,+,+,+)$.

The Einstein and gauge field equations derived from the action (\ref{ac})
are given by 
\begin{eqnarray}
R_{\mu \nu } &=&-3\ell ^{-2}g_{\mu \nu }-2F_{\mu \rho }F_{\nu }{}^{\rho }+{
\frac{1}{2}}F_{\alpha \beta }F^{\alpha \beta }g_{\mu \nu },  \notag \\
d\star F &=&0\ .
\end{eqnarray}

We shall consider solutions admitting a non-zero Killing spinor $\epsilon $
satisfying the Killing spinor equation: 
\begin{equation}
D_{\mu }\epsilon \equiv \left( \partial _{\mu }+{\frac{1}{4}}\omega _{\mu
,\nu _{1}\nu _{2}}\Gamma ^{\nu _{1}\nu _{2}}+{\frac{1}{4}}F_{\nu _{1}\nu
_{2}}\Gamma ^{\nu _{1}\nu _{2}}\Gamma _{\mu }+{\frac{1}{2}}\ell ^{-1}\Gamma
_{\mu }-\ell ^{-1}A_{\mu }\right) \epsilon =0\ .  \label{kse}
\end{equation}
The Killing spinor $\epsilon $ is a Dirac spinor. Following \cite{lawson},
this spinor is written as a differential form. The analysis of the conditions obtained from the Killing spinor equations makes use
of spinorial geometry techniques developed in \cite{Gillard:2004xq},
whereby the spinor is simplified into a number of canonical forms, and
a linear system of equations is obtained which imposes conditions on the spin connection and fluxes.

For the spinors in this paper, we shall make use of the canonical spinors
determined in \cite{preonone}. We
summarize a number of useful results here for convenience.
Dirac spinors are written as complexified forms on $\mathbb{R}^{2}$, and
a generic spinor $\eta$ can therefore be written as

\begin{equation}
\eta =\lambda 1+\mu ^{i}e_{i}+\sigma e_{12}
\end{equation}
where $e_{1}$, $e_{2}$ are 1-forms on $\mathbb{R}^{2}$, and $i=1,2$; 
$e_{12}=e_{1}\wedge e_{2}$. $\lambda $, $\mu ^{i}$ and $\sigma $ are complex
functions. It will be particularly useful to work in a null basis, and set 
\begin{equation}
\Gamma _{+}=\sqrt{2}i_{e_{2}}\ ,\qquad \Gamma _{-}=\sqrt{2}e_{2}\wedge \
,\qquad \Gamma _{1}=\sqrt{2}i_{e_{1}}\ ,\qquad \Gamma _{\bar{1}}=\sqrt{2}%
e_{1}\wedge \text{ .}
\end{equation}
In this basis the metric is given by 
\begin{equation}
ds^{2}=2\left( \mathbf{e}^{+}\mathbf{e}^{-}+\mathbf{e}^{1}\mathbf{e}^{{\bar{1
}}}\ \right) .
\end{equation}
In addition, the action of $\Gamma _{5}=\Gamma _{+-1\bar{1}}$ is given by 
\begin{equation}
\Gamma _{5}1=1,\qquad \Gamma _{5}e_{12}=e_{12},\qquad \Gamma
_{5}e_{i}=-e_{i}\ .
\end{equation}

Using $Spin(3,1)$ gauge transformations described explicitly in \cite{preonone}, a spinor $\epsilon $
can be written as one of three possible simple canonical forms: 
\begin{equation}
\epsilon =e_{2}  \label{form1}
\end{equation}
or 
\begin{equation}
\epsilon =1+\mu ^{1}e_{1}  \label{form2}
\end{equation}
or 
\begin{equation}
\epsilon =1+\mu ^{2}e_{2}\ .  \label{form3}
\end{equation}
Note that by making use of a $Spin(3,1)$ transformation generated by $\Gamma
_{+-}$, combined with an appropriately chosen $U(1)$ gauge transformation of 
$A$ which together leave $1$ invariant, one can without loss of generality
take $|\mu ^{2}|=1$ in ({\ref{form3}}).

To proceed, we evaluate the Killing spinor equation ({\ref{kse}}) acting on
the spinor 
\begin{eqnarray}
\epsilon = \lambda 1+\mu^{i}e_{i} \ .
\end{eqnarray}
The resulting equations are summarized in Appendix A. We then consider the
three cases ({\ref{form1}}), ({\ref{form2}}) and ({\ref{form3}}) separately.

\section{Solutions with $\protect\epsilon=e_2$}

In order to analyse solutions with $\epsilon =e_{2}$, we evaluate the
equations in Appendix A with $\lambda =\mu ^{1}=0$ and $\mu ^{2}=1$. One
obtains 
\begin{equation}
F_{+-}+F_{1{\bar{1}}}+\ell ^{-1}=0
\end{equation}
and 
\begin{equation}
F_{+-}+F_{1{\bar{1}}}-\ell ^{-1}=0\ .
\end{equation}
It is clear that these equations admit no solution; hence there are no
supersymmetric solutions with Killing spinor $\epsilon =e_{2}$.

\section{Solutions with $\protect\epsilon =1+\protect\mu e_{1}$}

These solutions split into two cases, according as $|\mu|=1$ and $|\mu|\neq
1 $, which will be considered separately.

\subsection{Solutions with $|\protect\mu| \neq 1$}

On evaluating the equations in Appendix A with $\lambda=1$, $\mu^1=\mu$, $\mu^2=0$ one obtains the conditions:

\begin{equation}
\partial _{1}\mu =\partial _{+}\mu =0,\text{ \ \ \ \ }\partial _{\bar{1}}\mu
=\sqrt{2}\ell ^{-1}\left( |\mu |^{2}-1\right) ,  \label{muder}
\end{equation}
\begin{eqnarray}
\omega _{+1} &=&0,  \notag \\
\omega _{1{\bar{1}}} &=&{\frac{1}{1-|\mu |^{2}}}(\bar{\mu}\partial _{-}\mu
-\mu \partial _{-}\bar{\mu})\mathbf{e}^{-}+\sqrt{2}\ell ^{-1}\left( \mu 
\mathbf{e}^{1}-\bar{\mu}\mathbf{e}^{{\bar{1}}}\right) ,  \label{sc}
\end{eqnarray}
\begin{equation}
\ell ^{-1}A=\frac{1}{2}d\log (1-|\mu |^{2})-\frac{1}{2}\omega _{+-},
\end{equation}
\begin{equation}
F=\frac{1}{\sqrt{2}\left( 1-|\mu |^{2}\right) }\mathbf{e}^{-}\wedge \left(
\partial _{-}\mu \mathbf{e}^{1}+\partial _{-}\bar{\mu}\mathbf{e}^{\bar{1}}+\sqrt{2}\ell ^{-1}\left( 1-|\mu |^{2}\right) \mathbf{e}^{+}\right) \text{ }.
\end{equation}

Using the conditions (\ref{sc}), we observe first that $\mathbf{e}^{-}$ is
hypersurface orthogonal and as such we introduce a co-ordinate $u$ and
function $L$ such that 
\begin{equation}
\mathbf{e}^{-}=Ldu\ .
\end{equation}
The conditions on the gauge potential $A$ imply that 
\begin{equation}
\ell ^{-1}A={\frac{1}{2}}d\log (1-|\mu |^{2})+{\frac{1}{2}}d\log L+Pdu
\end{equation}
for some function $P$. By making an appropriate $Spin(3,1)$ transformation
generated by $\Gamma _{+-}$ together with a $A$-gauge transformation leaving
the spinor $1+\mu e^{1}$ invariant, one can work in a gauge for which 
\begin{eqnarray}
\ell ^{-1}A &=&-{\frac{1}{2}}\left( \omega _{-,+-}+{\frac{\partial _{-}|\mu
|^{2}}{(1-|\mu |^{2})}}\right) \mathbf{e}^{-},  \notag \\
\omega _{+,+-} &=&0,  \notag \\
\omega _{{1},+-} &=&\sqrt{2}\ell ^{-1}\mu .
\end{eqnarray}
In this gauge, we then find 
\begin{equation}
d\mathbf{e}^{-}=-d\log (1-|\mu |^{2})\wedge \mathbf{e}^{-}
\end{equation}
and hence it is most convenient to introduce a local co-ordinate $u$ such
that 
\begin{equation}
\mathbf{e}^{-}={\frac{1}{1-|\mu |^{2}}}du\ .
\end{equation}
Next, the exterior derivative of $\mathbf{e}^{1}$ when restricted to
hypersurfaces of constant $u,$ gives 
\begin{equation}
{\hat{d}}\mathbf{e}^{1}=-{\hat{d}}\log (1-|\mu |^{2})\wedge \mathbf{e}^{1}
\end{equation}
where ${\hat{d}}$ denotes the restriction of the exterior derivative to $%
u=const.$. This implies that one can introduce a complex co-ordinate $z$
such that 
\begin{equation}
\mathbf{e}^{1}={\frac{1}{1-|\mu |^{2}}}(dz+\varrho du)  \label{basenull1}
\end{equation}
for $\varrho \in \mathbb{C}$. We can simplify the metric further by
performing the $Spin(3,1)$ gauge transformation generated by $\beta \Gamma
_{+1}+\bar{\beta}\Gamma _{+\bar{1}}$, for $\beta \in \mathbb{C}$, and which
leaves the Killing spinor $1+\mu e_{1}$ invariant. This gauge transformation
induces no change on $\mathbf{e}^{-}$ but sends $\mathbf{e}^{1}$ to $\mathbf{e}^{1}-2\beta \mathbf{e}^{-},$ therefore with an appropriate choice of $\beta $, one can, without loss of generality, set $\varrho =0$ in ({\ref{basenull1}}).

Finally we introduce the local co-ordinate $v$ such that the vector field
dual to $\mathbf{e}^{-}$ is ${\frac{\partial }{\partial v}}$, and we write 
\begin{equation}
\mathbf{e}^{+}=dv+\mathcal{H}du+\mathcal{G}dz+{\bar{\mathcal{G}}}d{\bar{z},}
\notag  \label{simplerbasis1}
\end{equation}
where $\mathcal{H}$ is a real function, $\mathcal{G}$ is a complex function,
and $\mu $ is independent of the coordinate $v$.

In terms of the introduced coordinates, the conditions ({\ref{muder}}) imply
that 
\begin{equation}
{d\mu }=-\sqrt{2}\ell ^{-1}d{\bar{z}+\frac{\partial _{-}\mu }{1-|\mu |^{2}}du}\text{ }.
\end{equation}
Performing a a change in co-ordinates of the form ${\bar{z}}^{\prime }={\bar{z}}+\psi (u)$ together with an appropriate $Spin(3,1)$ transformation
generated by $\beta \Gamma _{+1}+\bar{\beta}\Gamma _{+\bar{1}}$, one can set 
$\partial _{-}\mu =0,$ and thus 
\begin{equation}
\mu =-\sqrt{2}\ell ^{-1}{\bar{z}}\ .
\end{equation}
To proceed, consider the conditions $\omega _{-,+1}=\omega _{-,1\bar{1}}=0$
on the geometry. It is straightforward to show that these imply that 
\begin{equation}
{\mathcal{G}}={\frac{2\ell ^{-2}v\bar{z}}{1-2\ell ^{-2}z{\bar{z}}}}+\phi 
\end{equation}
where $\phi (u,z,{\bar{z}})$ is a complex function satisfying 
\begin{equation}
\partial _{z}\left( {\frac{\bar{\phi}}{1-2\ell ^{-2}z\bar{z}}}\right)
-\partial _{\bar{z}}\left( {\frac{\phi }{1-2\ell ^{-2}z\bar{z}}}\right) =0\ .
\end{equation}
The Bianchi identity $F=dA$ gives the conditions 
\begin{equation}
{\frac{\partial ^{2}\mathcal{H}}{\partial v^{2}}}=-{\frac{2\ell ^{-2}}{
1-2\ell ^{-2}z{\bar{z}}}},\qquad {\frac{\partial ^{2}\mathcal{H}}{\partial
z\partial v}}=-{\frac{2\ell ^{-2}}{1-2\ell ^{-2}z\bar{z}}}\bigg({\frac{2\ell
^{-2}v\bar{z}}{1-2\ell ^{-2}z{\bar{z}}}}+\phi \bigg)\ .
\end{equation}
These can be solved to find 
\begin{equation}
\mathcal{H}=-{\frac{\ell ^{-2}v^{2}}{1-2\ell ^{-2}z{\bar{z}}}}+\Theta
_{1}v+\Theta _{2}
\end{equation}
where $\Theta _{1},\Theta _{2}$ do not depend on $v$, and 
\begin{equation}
\phi =-{\frac{1}{2}}\ell ^{2}(1-2\ell ^{-2}z{\bar{z}}){\frac{\partial \Theta
_{1}}{\partial z}}\ .
\end{equation}
One can simplify the solution considerably by making the co-ordinate
transformation 
\begin{equation}
v=(1-2\ell ^{-2}z{\bar{z}})\left( v^{\prime }+{\frac{1}{2}}\ell ^{2}\Theta
_{1}\right) \ .
\end{equation}
On dropping the prime on $v^{\prime }$ the solution can then be written as 
\begin{equation}
ds^{2}=2du\left( dv+\left( -\ell ^{-2}v^{2}+\Psi \right) du\right) +{\frac{2}{(1-2\ell ^{-2}z{\bar{z}})^{2}}}dzd{\bar{z}}
\end{equation}
with 
\begin{equation}
F=-\ell ^{-1}dv\wedge du\ .
\end{equation}
The Einstein equations impose a further condition that $\Psi =\Psi (u,z,{\bar{z}})$ appearing in the metric is harmonic on $\mathbb{R}^{2}$: 
\begin{equation}
{\frac{\partial ^{2}\Psi }{\partial z\partial {\bar{z}}}}=0\ .
\end{equation}
Finally we note that no extra conditions arise from Maxwell equation.

\subsection{Solutions with $|\protect\mu|=1$}

On setting $\mu =e^{i\theta }$, the linear system in Appendix A implies that 
\begin{equation}
d\theta =0
\end{equation}
and hence we can use a gauge transformation generated by $i\Gamma _{1\bar{1}} $ to set, without loss of generality, $\mu =1$. It will furthermore be
useful to make a $U(1)$ gauge transformation, and a compensatory $Spin(1,3)$
gauge transformation generated by $\Gamma _{+-}$, which when combined leave
the spinor $1+e_{1}$ invariant, to set 
\begin{equation}
A_{+}=0.
\end{equation}
Then the conditions on the spin connection are 
\begin{equation}
\omega _{+,+1}=\omega _{1,+\bar{1}}=\omega _{+,1\bar{1}}=\omega _{+,+-}=0
\end{equation}
and 
\begin{equation}
\sqrt{2}\ell ^{-1}+\omega _{-,+1}-\omega _{1,1\bar{1}}=0
\end{equation}
and the conditions involving the gauge fields are 
\begin{equation}
\ell ^{-1}A_{1}={\frac{1}{2}}\left( \omega _{1,1\bar{1}}-\omega
_{1,+-}\right)
\end{equation}
with 
\begin{eqnarray}
F_{+1} &=&-{\frac{1}{\sqrt{2}}}\omega _{1,+1}, \\
F_{-1} &=&{\frac{1}{2\sqrt{2}}}\left( -\omega _{-,+-}+\omega _{-,1\bar{1}}-2\ell ^{-1}A_{-}\right) ,  \label{fm} \\
F_{+-} &=&-\ell ^{-1}-{\frac{1}{\sqrt{2}}}\left( \omega _{-,+1}+\omega _{-,+\bar{1}}\right) , \\
F_{1\bar{1}} &=&-{\frac{1}{\sqrt{2}}}\left( \omega _{-,+1}-\omega _{-,+\bar{1}}\right) \text{ }.
\end{eqnarray}
These conditions imply that

\begin{equation}
d\mathbf{e}^{-}=\mathbf{e}^{-}\wedge \bigg(\sqrt{2}\ell ^{-1}(\mathbf{e}^{1}+
\mathbf{e}^{\bar{1}})-2\ell ^{-1}A\bigg)\text{ }.  \label{hso1}
\end{equation}
In particular, as $\mathbf{e}^{-}\wedge d\mathbf{e}^{-}=0$, $\mathbf{e}^{-}$
is hypersurface orthogonal, so one can introduce a co-ordinate $u$ and
function $H$ such that 
\begin{equation}
\mathbf{e}^{-}=Hdu
\end{equation}
and it follows from ({\ref{hso1}}) that 
\begin{equation}
2\ell ^{-1}A=\sqrt{2}\ell ^{-1}(\mathbf{e}^{1}+\mathbf{e}^{\bar{1}})+H^{-1}dH+Qdu
\end{equation}
for some function $Q$. If the vector field dual to $\mathbf{e}^{-}$ is $\mathbf{e}^{-}={\frac{\partial }{\partial v}}$, then as $A_{+}=0$ it follows
that $\partial _{v}H=0$.

Next, consider a $U(1)$ gauge transformation $A=A^{\prime }+{\frac{\ell }{2}}H^{-1}dH$, which preserves the gauge $A_{+}=0$, together with a compensatory 
$Spin(3,1)$ gauge transformation generated by $\Gamma _{+-}$, which together
leave the spinor $1+e_{1}$ invariant. Using these, one can work in a gauge
for which 
\begin{equation}
2\ell ^{-1}A=\sqrt{2}\ell ^{-1}(\mathbf{e}^{1}+\mathbf{e}^{\bar{1}})+Qdu,\qquad {\mathbf{e}^{-}}=du\text{ }.  \label{gpp1}
\end{equation}

It is also useful to consider various spinor bilinears constructed from $\epsilon =1+e_{1}$. We recall from \cite{preonone} that there is a
non-degenerate $Spin(3,1)$ invariant inner product on Dirac spinors given by 
\begin{equation}
{\cal{B}}(\eta, \epsilon) = \langle B* \eta, \epsilon \rangle
\end{equation}
where
\begin{equation}
B.1 = -e_{12}, \qquad B e_{12}=1, \qquad B e_i = -\epsilon_i{}^j e_j \ .
\end{equation}
In particular, for the spinor $\epsilon =1+e_{1},$ we obtain 
\begin{equation}
{\cal{B}}(\epsilon ,\epsilon )={\cal{B}}(\epsilon ,\Gamma _{5}\epsilon )={\cal{B}}(\epsilon ,\Gamma
_{\mu }\epsilon )=0\text{ }.
\end{equation}
However, there is a non-zero 1-form spinor bilinear 
\begin{equation}
W_{\mu }={\cal{B}}(\epsilon ,\Gamma _{5}\Gamma _{\mu }\epsilon )
\end{equation}
which in the basis we have chosen gives 
\begin{equation}
W=2\sqrt{2}\mathbf{e}^{-}.
\end{equation}
There is also a non-zero 2-form spinor bilinear 
\begin{equation}
\chi _{\mu \nu }={\cal{B}}(\epsilon ,\Gamma _{5}\Gamma _{\mu \nu }\epsilon )
\end{equation}
which is 
\begin{equation}
\chi =2\mathbf{e}^{-}\wedge (\mathbf{e}^{1}+\mathbf{e}^{\bar{1}})\text{ }.
\end{equation}

One finds that the Killing spinor equation implies that 
\begin{equation}
\nabla _{\nu }W_{\mu }=2\ell ^{-1}A_{\nu }W_{\mu }-\ell ^{-1}\chi _{\mu \nu
}-F_{\mu \lambda }\chi ^{\lambda }{}_{\nu }-F_{\nu \lambda }\chi ^{\lambda
}{}_{\mu }-{\frac{1}{2}}F_{\lambda _{1}\lambda _{2}}\chi ^{\lambda
_{1}\lambda _{2}}\delta _{\mu \nu }  \label{bl1}
\end{equation}
and 
\begin{equation}
\nabla _{\nu }\chi _{\mu _{1}\mu _{2}}=2\ell ^{-1}A_{\nu }\chi _{\mu _{1}\mu
_{2}}+W_{\nu }F_{\mu _{1}\mu _{2}}+\delta _{\nu \mu _{1}}F_{\mu _{2}\lambda
}W^{\lambda }-\delta _{\nu \mu _{2}}F_{\mu _{1}\lambda }W^{\lambda }
\label{bl2}
\end{equation}
and we work in a basis for which 
\begin{equation}
dW=0,\qquad \chi =W\wedge A
\end{equation}
\begin{equation}
d\chi =-W\wedge F\text{ }.
\end{equation}
However, ({\ref{bl2}}) implies that 
\begin{equation}
d\chi =2\ell ^{-1}A\wedge \chi +W\wedge F\text{ }.
\end{equation}
It follows that 
\begin{equation}
W\wedge F=0\text{ }.
\end{equation}
This implies that 
\begin{equation}
F_{1{\bar{1}}}=0,\qquad F_{+1}=0\text{ }.  \label{mcond1}
\end{equation}

Furthermore, we remark that from the expression for the gauge potential
given by ({\ref{gpp1}}) we find that 
\begin{equation}
2\ell ^{-1}F_{+-}=\sqrt{2}\ell ^{-1}d(\mathbf{e}^{1}+\mathbf{e}^{\bar{1}})_{+-}+\partial _{+}Q\text{ }.
\end{equation}
However, a $Spin(3,1)$ gauge transformation generated by $\Gamma
_{+1}+\Gamma _{+\bar{1}}$, which leaves the spinor $1+e_{1}$ invariant, can
be used to set the value of $Q$ to any value, and so without loss of
generality $Q$ can be chosen such that 
\begin{equation}
F_{+-}=0\text{ }.  \label{mcond2}
\end{equation}

Given the conditions ({\ref{mcond1}}) and ({\ref{mcond2}}) on the Maxwell
field-strength components, the conditions on the spin connection can be
rewritten as 
\begin{equation}
d\mathbf{e}^{-}=0,
\end{equation}
and 
\begin{equation}
(d\mathbf{e}^{1})_{+1}=(d\mathbf{e}^{\bar{1}})_{+1}=0,  \label{de1cond}
\end{equation}
and 
\begin{equation}
\omega _{1,1\bar{1}}={\frac{1}{\sqrt{2}}}\ell ^{-1},\text{ \ \ }\omega
_{-,+1}=-{\frac{1}{\sqrt{2}}}\ell ^{-1}\text{ }.  \label{geoextra1}
\end{equation}
The remaining content of the Bianchi identity from ({\ref{gpp1}}) is given
by 
\begin{equation}
\sqrt{2}\ell ^{-1}d(\mathbf{e}^{1}+\mathbf{e}^{\bar{1}})_{+-}+\partial
_{+}Q=0  \label{bbian1}
\end{equation}
and 
\begin{equation}
2\ell ^{-1}F_{-1}=\sqrt{2}\ell ^{-1}d(\mathbf{e}^{1}+\mathbf{e}^{\bar{1}})_{-1}-\partial _{1}Q\text{ }.  \label{bbian2}
\end{equation}

We proceed by introducing co-ordinates. We have already introduced
co-ordinates $u,v$ such that $\mathbf{e}^{-}=du$ and the vector field dual
to $\mathbf{e}^{-}$ is ${\frac{\partial }{\partial v}}$. As $d(\mathbf{e}^{1}+\mathbf{e}^{\bar{1}})$ vanishes when restricted to surfaces of constant 
$u$, it follows that we can introduce a real co-ordinate $x$, and a real
function $P$ such that 
\begin{equation}
\mathbf{e}^{1}+\mathbf{e}^{\bar{1}}=dx+Pdu  \label{coordeq1}
\end{equation}
and moreover, we can also introduce another real co-ordinate $y$ such that 
\begin{equation}
i(\mathbf{e}^{1}-\mathbf{e}^{\bar{1}})=fdx+gdy+Ldu  \label{coordeq2}
\end{equation}
for real functions $f,g,L$. We remark that $L$ can be set to zero without
loss of generality, by making use of a gauge transformation generated by $i\Gamma _{+1}-i\Gamma _{+\bar{1}}$, which leaves the spinor $1+e_{1}$
invariant, and also does not induce any change to $\mathbf{e}^{1}+\mathbf{e}^{\bar{1}}$, so we take 
\begin{equation}
i(\mathbf{e}^{1}-\mathbf{e}^{\bar{1}})=fdx+gdy\text{ }.  \label{coordeq2b}
\end{equation}
The remaining basis element is then given by 
\begin{equation}
\mathbf{e}^{+}=dv+\Upsilon du+hdx+Sdy
\end{equation}
for real functions $\Upsilon ,h,S$. To proceed, we consider the condition ({\ref{de1cond}}). This implies that 
\begin{equation}
\partial _{v}g=\partial _{v}f=0\text{ }.
\end{equation}

It follows that we can make a $v$-independent shift in $y$ which preserves
the form of $\mathbf{e}^{+}$ and can be chosen in order to set $f=0$ in ({\ref{coordeq2b}}). Such a transformation re-introduces a $du$ term in this
expression, which can then be again eliminated by use of a gauge
transformation generated by $i\Gamma _{+1}-i\Gamma _{+\bar{1}}$. So, without
loss of generality, we set $f=0$ in ({\ref{coordeq2b}}), and $g$ is
independent of $v$.

Next consider the imaginary part of the second condition in ({\ref{geoextra1}}). On evaluating the appropriate components of the spin connection this
implies that 
\begin{equation}
\partial _{v}S=0
\end{equation}
and so by making an appropriately chosen $v$-independent shift in the $v$
co-ordinate we can without loss of generality also set $S=0$.

In addition, the real part of the second geometric condition in ({\ref{geoextra1}}) can be written as 
\begin{equation}
\partial _{v}\bigg(h+{\frac{1}{2}}P\bigg)=\sqrt{2}\ell ^{-1}\text{ }.
\end{equation}
The Bianchi identity ({\ref{bbian1}}) can be rewritten as 
\begin{equation}
\partial _{v}\bigg(Q+\sqrt{2}\ell ^{-1}P\bigg)=0
\end{equation}
whereas the real part of the second Bianchi identity ({\ref{bbian2}})
together with ({\ref{fm}}) imply that 
\begin{equation}
-\partial _{x}(Q+\sqrt{2}\ell ^{-1}P)={\frac{1}{\sqrt{2}}}\ell ^{-1}\bigg(\partial _{v}\Upsilon -P\partial _{v}H -Q \bigg)
\end{equation}
and the imaginary part implies 
\begin{equation}
\partial _{y}(Q+\sqrt{2}\ell ^{-1}P)={\frac{1}{\sqrt{2}}}\partial _{y}\bigg(
h+{\frac{1}{2}}P\bigg)\text{ }.
\end{equation}

In writing the metric and Maxwell fields, it will be useful to set 
\begin{equation}
\Upsilon +{\frac{1}{4}}P^{2}=\Theta ,\quad Q+\sqrt{2}\ell ^{-1}P=\Psi ,\quad
h+{\frac{1}{2}}P=\tau
\end{equation}
so that the previous conditions can be rewritten as 
\begin{equation}
\partial _{v}\tau =\sqrt{2}\ell ^{-1},\quad \partial _{v}\Psi =0,\quad
\partial _{v}\Theta =\Psi -\sqrt{2}\ell \partial _{x}\Psi  \label{ggeo1}
\end{equation}
and 
\begin{equation}
\partial _{y}(\tau -\sqrt{2}\ell \Psi )=0  \label{ggeo2}
\end{equation}
and the metric and field strength are 
\begin{equation}
ds^{2}=2du\bigg(dv+\Theta du+\tau dx\bigg)+{\frac{1}{2}}dx^{2}+{\frac{1}{2}}
g^{2}dy^{2}
\end{equation}
and 
\begin{equation}
F={\frac{\ell }{2}}d\Psi \wedge du\text{ }.
\end{equation}
In fact, ({\ref{ggeo1}}) and ({\ref{ggeo2}}) can be integrated to obtain 
\begin{equation}
\Theta ={\mathcal{F}}(u,x,y)+v(\Psi -\sqrt{2}\ell \partial _{x}\Psi )
\end{equation}
and 
\begin{equation}
\tau =\sqrt{2}\ell ^{-1}v+\sqrt{2}\ell \Psi +\phi (u,x)\text{ .}
\end{equation}
By making a co-ordinate transformation of the form 
\begin{equation}
v=v^{\prime }+{\mathcal{H}}(u,x)
\end{equation}
for an appropriately chosen function ${\mathcal{H}}$, we can set $\phi
(u,x)=0$ without loss of generality. The metric is therefore given by 
\begin{eqnarray}
ds^{2} &=&2du\bigg(dv+\big({\mathcal{F}}+v(\Psi -\sqrt{2}\ell \partial
_{x}\Psi )\big)du+(\sqrt{2}\ell ^{-1}v+\sqrt{2}\ell \Psi )dx\bigg)  \notag
\label{spmet1} \\
&+&{\frac{1}{2}}dx^{2}+{\frac{1}{2}}g^{2}dy^{2}
\end{eqnarray}
where the prime on $v^{\prime }$ has been dropped.

It remains to consider the function $g$ appearing in the metric. The first
geometric condition in ({\ref{geoextra1}}) can then be written as 
\begin{equation}
g^{-1}\partial _{x}g=-{\frac{1}{\sqrt{2}}}\ell ^{-1}
\end{equation}
so 
\begin{equation}
g=K(u,y)e^{-{\frac{1}{\sqrt{2}}}\ell ^{-1}x}\text{ }.
\end{equation}
It is possible to make a co-ordinate transformation 
\begin{equation}
Kdy+\alpha (u,y)du=dy^{\prime },\qquad v=v^{\prime }-{\frac{1}{2}}e^{-{\frac{1}{\sqrt{2}}}\ell ^{-1}x}\Lambda
\end{equation}
such that $\partial _{y}\alpha -\partial _{u}K=0$, and $\partial _{y}\Lambda
=\alpha $, and work in co-ordinates such that $K=1$ without loss of
generality. So the metric is given by ({\ref{spmet1}}) with 
\begin{equation}
g=e^{-{\frac{1}{\sqrt{2}}}\ell ^{-1}x}\text{ }.
\end{equation}

We also consider the Maxwell equations obtained by computing 
\begin{equation}
2\ell ^{-1}\star F=du\wedge \bigg(e^{{\frac{1}{\sqrt{2}}}\ell
^{-1}x}\partial _{y}\Psi dx-e^{-{\frac{1}{\sqrt{2}}}\ell ^{-1}x}\partial
_{x}\Psi dy\bigg)
\end{equation}
On requiring $d\star F=0$, we find the condition 
\begin{equation}
{\frac{\partial ^{2}\Psi }{\partial x^{2}}}-{\frac{1}{\sqrt{2}}}\ell ^{-1}{\frac{\partial \Psi }{\partial x}}+e^{\sqrt{2}\ell ^{-1}x}{\frac{\partial
^{2}\Psi }{\partial y^{2}}}=0  \label{mxeq1}
\end{equation}
or equivalently 
\begin{equation}
\Box _{2}\Psi =0
\end{equation}
where $\Box _{2}$ is the Laplacian on $H^{2}$, equipped with metric 
\begin{equation}
ds_{2}^{2}={\frac{1}{2}}\left( dx^{2}+e^{-\sqrt{2}\ell ^{-1}x}dy^{2}\right) 
\text{ }.
\end{equation}

The remaining content of the Einstein field equations is given by the $v$-independent part of the $uu$ component, which implies that 
\begin{eqnarray}
\label{eeinspc}
{\partial^2 {\cal{F}} \over \partial x^2}
+  e^{\sqrt{2} \ell^{-1}x} {\partial^2 {\cal{F}} \over \partial y^2}
+{1 \over \sqrt{2}} \ell^{-1} {\partial {\cal{F}} \over \partial x}
- \ell^{-2} {\cal{F}}
&=& -{\partial \Psi \over \partial u}
+4 \sqrt{2} \ell \Psi {\partial \Psi \over \partial x} - \Psi^2
\nonumber \\
&-&{3 \over 2} \ell^2 \bigg({\partial \Psi \over \partial x} \bigg)^2
+\ell \sqrt{2} {\partial^2 \Psi \over \partial x \partial u}
\nonumber \\
&+&{5 \over 2} \ell^2 e^{\sqrt{2} \ell^{-1}x}\bigg({\partial \Psi \over \partial y} \bigg)^2
-4 \ell^2 \Psi {\partial^2 \Psi \over \partial x^2} \ .
\nonumber \\
\end{eqnarray}

So for this class of solution, the metric is given in terms of two functions 
$\Psi (u,x,y)$ and ${\mathcal{F}}(u,x,y)$ via 
\begin{eqnarray}
ds^{2} &=&2du\bigg(dv+\big({\mathcal{F}}+v(\Psi -\sqrt{2}\ell \partial
_{x}\Psi )\big)du+\sqrt{2}\big(\ell ^{-1}v+\ell \Psi \big)dx\bigg)
+ds_{2}^{2}.  \notag \\
&&
\end{eqnarray}
where 
\begin{equation}
ds_{2}^{2}={\frac{1}{2}}\left( dx^{2}+e^{-\sqrt{2}\ell ^{-1}x}dy^{2}\right)
\end{equation}
is the metric on $H^{2}$, and 
\begin{equation}
F={\frac{\ell }{2}}d\Psi \wedge du\text{ }.
\end{equation}
The Maxwell field equations imply that $\Psi $ is harmonic on $H^{2}$, and
the Einstein equations reduce to the condition ({\ref{eeinspc}}).
On setting $k={\partial \over \partial v}$, the Maxwell field strength satisfies
\begin{equation}
F_{\mu \lambda} F_\nu{}^\lambda = {\ell^2 \over 2} \bigg(\big({\partial \Psi \over \partial x}\big)^2 + e^{\sqrt{2} \ell^{-1} x}
\big({\partial \Psi \over \partial y} \big)^2 \bigg) k_\mu k_\nu \ .
\end{equation}
It follows that the geometry is a type III Kundt solution, coupled to a pure radiation field. Such solutions were constructed in
\cite{kundt1, kundt2} with vanishing cosmological constant. 
Here there is a non-vanishing cosmological constant, and the solution is a subcase of the classification of such solutions constructed in \cite{Griffiths}.

\section{Solutions with $\protect\epsilon =1+ e^{i \protect\theta} e_{2}$}

On evaluating the equations in Appendix A with $\lambda=1$, $\mu^1=0$, $\mu^2 = e^{i \theta}$, one obtains the components of the gauge field
strength as:

\begin{eqnarray}
F_{+-} &=&-\sqrt{2}\big(\cos \theta \omega _{+,+-}+\partial _{+}\theta \sin
\theta \big)+\ell ^{-1},  \notag  \label{fluxc1} \\
F_{1{\bar{1}}} &=&-i\sqrt{2}\big(-\sin \theta \omega _{+,+-}+\cos \theta
\partial _{+}\theta \big),  \notag \\
F_{-1} &=&-\frac{1}{\sqrt{2}}e^{i\theta }\omega _{-,-1},  \notag \\
F_{+1} &=&-\frac{1}{\sqrt{2}}e^{-i\theta }\omega _{+,+1}\ .
\end{eqnarray}
The components of the gauge potential are given by: 
\begin{equation}
\ell ^{-1}A_{-}=-{\frac{1}{2}}\omega _{-,+-},\ \ \ \ell ^{-1}A_{1}=\frac{1}{2}\left( i\partial _{1}\theta -\omega _{1,1{\bar{1}}}\right) ,\ \ \ \ell
^{-1}A_{+}=\frac{1}{2}\omega _{+,+-}\ .  \label{fluxc2}
\end{equation}
The geometric constraints are given by 
\begin{eqnarray}
\omega _{+,+-}-\omega _{-,+-} &=&\sqrt{2}\ell ^{-1}\cos \theta ,  \notag \\
\partial _{-}\theta +\partial _{+}\theta &=&\sqrt{2}\ell ^{-1}\sin \theta , 
\notag \\
\omega _{{\bar{1}},1{\bar{1}}} &=&2i\partial _{\bar{1}}\theta -\omega _{+,+\bar{1}}=-\omega _{-,-\bar{1}},  \notag \\
\omega _{{\bar{1}},+1} &=&-\omega _{+,+-}-i\partial _{+}\theta +\sqrt{2}e^{i\theta }\ell ^{-1},  \notag \\
\omega _{{\bar{1}},-1} &=&\omega _{+,+-}-i\partial _{+}\theta ,  \notag \\
\omega _{1,+-} &=&-i\partial _{1}\theta ,\ \ \omega _{+,1{\bar{1}}}=2i\partial _{+}\theta ,  \notag \\
\omega _{-,1{\bar{1}}} &=&\omega _{+,-1}=\omega _{-,+1}=\omega
_{1,+1}=\omega _{1,-1}=0\ .
\end{eqnarray}
Thus we can write \bigskip 
\begin{eqnarray}
d\mathbf{e}^{1} &=&\left( i\partial _{+}\theta +\omega _{+,+-}-\sqrt{2}\ell
^{-1}e^{-i\theta }\right) \mathbf{e}^{+}\wedge \mathbf{e}^{1}  \notag
\label{extder1} \\
&+&\left( -\omega _{+,+-}-i\partial _{+}\theta \right) \mathbf{e}^{-}\wedge 
\mathbf{e}^{1}+\left( -2i\partial _{\bar{1}}\theta +\omega _{+,+\bar{1}
}\right) \mathbf{e}^{1}\wedge \mathbf{e}^{\bar{1}},  \notag \\
d\mathbf{e}^{+} &=&-\omega _{-,+-}\mathbf{e}^{+}\wedge \mathbf{e}^{-}-\omega
_{-,-1}\mathbf{e}^{-}\wedge \mathbf{e}^{1}-\omega _{-,-{\bar{1}}}\mathbf{e}
^{-}\wedge \mathbf{e}^{{\bar{1}}}  \notag \\
&+&i\partial _{1}\theta \mathbf{e}^{+}\wedge \mathbf{e}^{1}-i\partial _{\bar{1}}\theta \mathbf{e}^{+}\wedge \mathbf{e}^{{\bar{1}}}-2i\partial _{+}\theta 
\mathbf{e}^{1}\wedge \mathbf{e}^{\bar{1}},  \notag \\
d\mathbf{e}^{-} &=&-\omega _{+,+-}\mathbf{e}^{+}\wedge \mathbf{e}^{-}-i\partial _{1}\theta \mathbf{e}^{-}\wedge \mathbf{e}^{1}+i\partial _{\bar{1}}\theta \mathbf{e}^{-}\wedge \mathbf{e}^{{\bar{1}}}  \notag \\
&-&\omega _{+,+1}\mathbf{e}^{+}\wedge \mathbf{e}^{1}-\omega _{+,+\bar{1}}\mathbf{e}^{-}\wedge \mathbf{e}^{\mathbf{1}}+2i\left( -\partial _{+}\theta +
\sqrt{2}\ell ^{-1}\sin \theta \right) \mathbf{e}^{1}\wedge \mathbf{e}^{\bar{1}}\ .
\end{eqnarray}

\subsection{Solutions with $\sin \protect\theta \neq 0$}

For these solutions, it is convenient to define the 1-form 
\begin{equation}
V={\frac{1}{\sin \theta }}(\mathbf{e}^{+}+\mathbf{e}^{-})
\end{equation}
and introduce a local co-ordinate $\tau $ such that $V={\frac{\partial }{\partial \tau }}$.

It is straightforward to see that the supersymmetry constraints imply that 
\begin{equation}
{\frac{\partial \theta }{\partial \tau }}=\sqrt{2}\ell ^{-1}
\end{equation}
and furthermore 
\begin{eqnarray}
\mathcal{L}_{V}\mathbf{e}^{1} &=&-{\frac{\sqrt{2}\ell ^{-1}e^{-i\theta }}{\sin \theta }}\mathbf{e}^{1},  \notag \\
\mathcal{L}_{V}(\mathbf{e}^{+}-\mathbf{e}^{-}) &=&-\sqrt{2}\ell ^{-1}\cot
\theta (\mathbf{e}^{+}-\mathbf{e}^{-})\ .
\end{eqnarray}
These constraints imply that one can write 
\begin{eqnarray}
\mathbf{e}^{1} &=&\big(1-i\cot \theta \big){\hat{\mathbf{e}}}^{1},  \notag \\
\mathbf{e}^{+}-\mathbf{e}^{-} &=&{\frac{\sqrt{2}}{\sin \theta }}{\hat{\mathbf{e}}}^{0},
\end{eqnarray}
where 
\begin{equation}
\mathcal{L}_{V}{\hat{\mathbf{e}}}^{1}=0,\qquad \mathcal{L}_{V}{\hat{\mathbf{e}}}^{0}=0\ .  \label{liederv1}
\end{equation}

Note, furthermore, that 
\begin{eqnarray}
d{\hat{\mathbf{e}}}^{1} &=&\bigg({\frac{\sqrt{2}}{\sin \theta }}\omega
_{+,+-}+{\frac{1}{\sqrt{2}}}{\frac{\cos \theta }{\sin ^{2}\theta }}(\partial
_{+}\theta -\partial _{-}\theta )  \notag  \label{based1} \\
&-&\ell ^{-1}\cot \theta +2i\ell ^{-1}\bigg){\hat{\mathbf{e}}}^{0}\wedge {\hat{\mathbf{e}}}^{1}  \notag \\
&+&{\frac{1}{\sin ^{2}\theta }}\bigg(-i\partial _{\bar{1}}\theta +i\sin
\theta e^{-i\theta }\omega _{+,+\bar{1}}\bigg){\hat{\mathbf{e}}}^{1}\wedge {\hat{\mathbf{e}}}^{\bar{1}}\ .
\end{eqnarray}
and 
\begin{eqnarray}
d\mathbf{e}^{2} &=&{\frac{1}{2\sin ^{2}\theta }}\bigg((e^{2i\theta }\omega
_{+,+1}-\omega _{-,-1}){\hat{\mathbf{e}}}^{1}+(e^{-2i\theta }\omega _{+,+\bar{1}}-\omega _{-,-\bar{1}}){\hat{\mathbf{e}}}^{\bar{1}}\bigg)\wedge {\hat{\mathbf{e}}}^{0}  \notag  \label{based2} \\
&-&2i\ell ^{-1}{\hat{\mathbf{e}}}^{1}\wedge {\hat{\mathbf{e}}}^{\bar{1}}
\text{ }.
\end{eqnarray}

On defining

\begin{eqnarray}
{\mathcal{B}} &=&{\frac{1}{2\sin ^{2}\theta }}\bigg(\big(e^{2i\theta }\omega
_{+,+1}-\omega _{-,-1}\big){\hat{\mathbf{e}}}^{1}+\big(e^{-2i\theta }\omega
_{+,+\bar{1}}-\omega _{-,-\bar{1}}\big){\hat{\mathbf{e}}}^{\bar{1}}\bigg) 
\notag  \label{bgt1} \\
&+&\bigg({\frac{\sqrt{2}}{\sin \theta }}\omega _{+,+-}-\ell ^{-1}\cot \theta
+{\frac{\cos \theta }{\sqrt{2}\sin ^{2}\theta }}(\partial _{+}\theta
-\partial _{-}\theta )\bigg){\hat{\mathbf{e}}}^{0},
\end{eqnarray}
the conditions ({\ref{based1}}) and ({\ref{based2}}) can be rewritten as 
\begin{eqnarray}
d{\hat{\mathbf{e}}}^{0} &=&{\mathcal{B}}\wedge {\hat{\mathbf{e}}}^{0}-2i\ell
^{-1}{\hat{\mathbf{e}}}^{1}\wedge {\hat{\mathbf{e}}}^{\bar{1}},  \notag
\label{based3} \\
d{\hat{\mathbf{e}}}^{1} &=&{\mathcal{B}}\wedge {\hat{\mathbf{e}}}^{1}-2i\ell
^{-1}{\hat{\mathbf{e}}}^{1}\wedge {\hat{\mathbf{e}}}^{0}\text{ }.
\end{eqnarray}

Noting then that the metric can be written as 
\begin{equation}
ds^{2}={\frac{1}{2}}(\mathbf{e}^{+}+\mathbf{e}^{-})^{2}+{\frac{1}{\sin
^{2}\theta }}\ ds_{LGT}^{2}
\end{equation}
where 
\begin{equation}
ds_{LGT}^{2}=-({\hat{\mathbf{e}}}^{0})^{2}+2{\hat{\mathbf{e}}}^{1}{\hat{\mathbf{e}}}^{\bar{1}}\text{ }.
\end{equation}
We note that the metric $ds_{LGT}^{2}$ on the Lorentzian 3-manifold $LGT$
does not depend on $\tau $, and moreover, the conditions ({\ref{based3}})
imply that $LGT$ admits a $\tau $-independent basis $\mathbf{E}^{i}$ for $i=1,2,3$ satisfying 
\begin{equation}
d\mathbf{E}^{i}={\mathcal{B}}\wedge \mathbf{E}^{i}+2\ell ^{-1}\star _{3}\mathbf{E}^{i}  \label{lgaudtod}
\end{equation}
where $\star _{3}$ denotes the Hodge dual on $LGT$ (in our conventions, the
volume form on $LGT$ is $i{\hat{\mathbf{e}}}^{1}\wedge {\hat{\mathbf{e}}}^{\bar{1}}\wedge {\hat{\mathbf{e}}}^{0}$). Note in particular that ({\ref{lgaudtod}}) implies that ${\mathcal{B}}$ must be independent of $\tau $,
and furthermore, must satisfy 
\begin{equation}
d{\mathcal{B}}=-2\ell ^{-1}\star _{3}{\mathcal{B}}  \label{intcond1}
\end{equation}
and in turn ({\ref{intcond1}}) implies that 
\begin{equation}
d\star _{3}{\mathcal{B}}=0\ .  \label{intcond2}
\end{equation}

The condition ({\ref{lgaudtod}}) implies that $LGT$ admits a \textit{%
Lorentzian Gauduchon-Tod structure}. Euclidean Gauduchon-Tod structures
arise in the context of 4-dimensional hyper-K\"{a}hler with torsion
manifolds which admit a tri-holomorphic isometry, and have been analysed in 
\cite{gtod} and \cite{mactod}. Such structures are also known to arise in
the context of both Euclidean and Lorentzian supergravity solutions \cite{ds, meessen}. Lorentzian Gauduchon-Tod
structures have also been found in gauged 4-dimensional
supergravity with signature $(+,+,-,-)$ \cite{Klemm2015}.

To proceed further, we next consider the constraints which ({\ref{extder1}})
impose on $\mathbf{e}^{+}+\mathbf{e}^{-}$. It will be convenient to write 
\begin{equation}
\mathbf{e}^{+}+\mathbf{e}^{-}={\frac{2}{\sin \theta }}(d\tau +\Omega )
\end{equation}
and to set 
\begin{equation}
\theta =\sqrt{2}\ell ^{-1}\tau +\Phi
\end{equation}
where $\Omega $ is a $\tau $-dependent 1-form on $LGT$ and $\Phi $ is a $\tau $-independent function. Then ({\ref{extder1}}) implies that 
\begin{equation}
{\mathcal{L}}_{V}\Omega -2\sqrt{2}\ell ^{-1}\cot \theta \ \Omega -{\mathcal{B}}+2\cot \theta \ d\Phi =0\ .
\end{equation}
This condition can be integrated up, and on changing co-ordinates from $\tau 
$ to $\theta $, one obtains 
\begin{equation}
\mathbf{e}^{+}+\mathbf{e}^{-}={\frac{\sqrt{2}\ell }{\sin \theta }}d\theta -\sqrt{2}\ell \cos \theta \ {\mathcal{B}}+2\sin \theta \ \psi
\end{equation}
where $\psi $ is a $\theta $-independent 1-form on $LGT$. Note that $\psi $
is defined in terms of the basis ${\hat{\mathbf{e}}}^{1},{\hat{\mathbf{e}}}^{\bar{1}},{\hat{\mathbf{e}}}^{0}$ as 
\begin{eqnarray}
\psi &=&{\frac{i\ell }{2\sqrt{2}\sin ^{2}\theta }}\bigg((\omega
_{-,-1}+e^{2i\theta }\omega _{+,+1}){\hat{\mathbf{e}}}^{1}-(\omega _{-,-\bar{1}}+e^{-2i\theta }\omega _{+,+\bar{1}}){\hat{\mathbf{e}}}^{\bar{1}}\bigg) 
\notag \\
&+&{\frac{1}{2}}\bigg({\frac{2\ell \cos \theta }{\sin ^{2}\theta }}\omega
_{+,+-}-{\frac{\sqrt{2}\cos ^{2}\theta }{\sin ^{2}\theta }}-{\frac{\ell }{\sin \theta }}(\partial _{+}\theta -\partial _{-}\theta )\bigg){\hat{\mathbf{e}}}^{0}.
\end{eqnarray}
The remaining content of ({\ref{extder1}}) imposes an additional condition
on $\psi $: 
\begin{equation}
d\psi +{\mathcal{B}}\wedge \psi +2\ell ^{-1}\star _{3}\psi =0\ .
\label{psicon1}
\end{equation}

It remains to consider the constraints on the fluxes. Note first that ({\ref{fluxc2}}) implies that 
\begin{equation}
A={\frac{\ell }{2}}\cot \theta d\theta -{\frac{\ell }{2}}\cos 2\theta \ {\mathcal{B}}+{\frac{1}{\sqrt{2}}}\sin 2\theta \ \psi \ .  \label{fluxpotx1}
\end{equation}
It is straightforward to show that on applying the exterior derivative to ({\ref{fluxpotx1}}), one obtains the components of the field strength given in
({\ref{fluxc1}}), with no further constraint. In order to evaluate the gauge
field equations, observe that the above conditions imply that the Hodge dual
of $F$ is given by 
\begin{eqnarray}
\star F &=&d\theta \wedge \bigg(\ell \cos 2\theta \ {\mathcal{B}}-\sqrt{2}
\sin 2\theta \ \psi \bigg)+\sqrt{2}\sin ^{2}\theta {\mathcal{B}}\wedge \psi 
\notag \\
&-&\star _{3}\bigg(\sin 2\theta \ {\mathcal{B}}+\sqrt{2}\ell ^{-1}\cos
2\theta \ \psi \bigg).
\end{eqnarray}
On making use of the previously obtained conditions, this expression can be
rewritten as 
\begin{equation}
\star F=d\bigg({\frac{\ell }{2}}\sin 2\theta \ {\mathcal{B}}-\sqrt{2}\sin
^{2}\theta \psi \bigg)-\sqrt{2}\ell ^{-1}\star _{3}\psi
\end{equation}
and we remark that the conditions ({\ref{intcond1}}) and ({\ref{psicon1}})
imply that $\star _{3}\psi $ is closed. Hence the gauge equations are
satisfied with no further conditions.

In order to examine the Einstein equations we follow the reasoning presented
(in the context of solutions of the anti-de-Sitter minimal gauged
supergravity) in Appendix E of \cite{klemm2003}. In particular, the
integrability conditions of the Killing spinor equation associated with a
pseudo-supersymmetric solution for which the Maxwell field strength $F$
satisfies the Bianchi identity and gauge field equations imply that 
\begin{equation}
E_{\mu \nu }\Gamma ^{\nu }\epsilon =0  \label{einint}
\end{equation}
where 
\begin{equation}
E_{\mu \nu }=R_{\mu \nu }+3\ell ^{-2}g_{\mu \nu }+2F_{\mu \rho }F_{\nu
}{}^{\rho }-{\frac{1}{2}}F_{\alpha \beta }F^{\alpha \beta }g_{\mu \nu }\ .
\end{equation}
Evaluating ({\ref{einint}}) acting on the Killing spinor $\epsilon
=1+e^{i\alpha }e_{2}$, one finds that all components of $E_{\mu \nu }$ are
constrained to vanish, i.e. the Einstein equations hold automatically.

To summarize, the solutions with Killing spinor $1+e^{i\theta }e_{2}$ and $\sin \theta \neq 0$ have metric 
\begin{equation}
ds^{2}=\bigg({\frac{\ell }{\sin \theta }}d\theta -\ell \cos \theta \ {%
\mathcal{B}}+\sqrt{2}\sin \theta \ \psi \bigg)^{2}+{\frac{1}{\sin ^{2}\theta 
}}ds_{LGT}^{2}
\end{equation}
where $ds_{LGT}^{2}$ is a $\theta $-independent metric on a 3-dimensional
Lorentzian manifold which has a Lorentzian Gauduchon-Tod structure. The
3-manifold $LGT$ admits a $\theta $-independent basis $\mathbf{E}^{i}$ and a 
$\theta $-independent 1-form ${\mathcal{B}}$ satisfying ({\ref{lgaudtod}})
(with associated integrability conditions ({\ref{intcond1}}) and ({\ref
{intcond2}})). $LGT$ also admits a $\theta $-independent 1-form $\psi $
satisfying ({\ref{psicon1}}). The flux is then given by 
\begin{equation}
F=d\bigg(-{\frac{\ell }{2}}\cos 2\theta \ {\mathcal{B}}+{\frac{1}{\sqrt{2}}}
\sin 2\theta \ \psi \bigg)\ .
\end{equation}
Examples of such solutions will be described in \cite{mdg}.

\subsection{Solutions with $\sin \protect\theta =0$}

If $\sin \theta =0$ then $\cos \theta = \pm 1$. However, as the spinor $%
1-e_1 $ is related to the spinor $1+e_1$ on making use of a $Pin$
transformation generated by $\Gamma_+ -\Gamma_-$, it is sufficient to
consider the spinor $\epsilon=1+e_1$ in this case.

\begin{eqnarray}
F &=&\left( -\sqrt{2}\omega _{+,+-}+\ell ^{-1}\right) \mathbf{e}^{+}\wedge 
\mathbf{e}^{-}-{\frac{1}{\sqrt{2}}}\left( \mathbf{e}^{+}+\mathbf{e}^{-}\right) \wedge \left( \omega _{+,+1}\mathbf{e}^{1}+\omega _{+,+\bar{1}}
\mathbf{e}^{\bar{1}}\right)  \notag  \label{gaugefs1} \\
&&
\end{eqnarray}
and 
\begin{equation}
\ell ^{-1}A={\frac{1}{2}}\omega _{+,+-}\mathbf{e}^{+}-{\frac{1}{2}}\omega
_{-,+-}\mathbf{e}^{-}-{\frac{1}{2}}\omega _{+,+1}\mathbf{e}^{1}-{\frac{1}{2}}
\omega _{+,+\bar{1}}\mathbf{e}^{\bar{1}}  \label{gaugepot1}
\end{equation}
and

\begin{eqnarray}
\omega _{+,+-}-\omega _{-,+-} &=&\sqrt{2}\ell ^{-1},  \notag \\
\hskip15mm\omega _{1,1{\bar{1}}} &=&\omega _{+,+1}=\omega _{-,-1},  \notag \\
\omega _{{\bar{1}},+1}+\omega _{+,+-} &=&\sqrt{2}\ell ^{-1},  \notag \\
\hskip15mm\omega _{{\bar{1}},-1} &=&\omega _{+,+-},  \notag \\
\hskip15mm\omega _{1,+-} &=&\omega _{+,1{\bar{1}}}=\omega _{-,1{\bar{1}}
}=\omega _{+,-1}=\omega _{-,+1}=\omega _{1,+1}=\omega _{1,-1}=0\ .
\end{eqnarray}

It follows that 
\begin{eqnarray}
d\mathbf{e}^{+} &=&-\omega _{-,+-}\mathbf{e}^{+}\wedge \mathbf{e}^{-}+\big(
\omega _{+,+1}\mathbf{e}^{1}+\omega _{+,+\bar{1}}\mathbf{e}^{\bar{1}}\big)
\wedge \mathbf{e}^{-},  \notag  \label{extalg} \\
d\mathbf{e}^{-} &=&-\omega _{+,+-}\mathbf{e}^{+}\wedge \mathbf{e}^{-}+\big(
\omega _{+,+1}\mathbf{e}^{1}+\omega _{+,+\bar{1}}\mathbf{e}^{\bar{1}}\big)
\wedge \mathbf{e}^{+},  \notag \\
d\mathbf{e}^{1} &=&\bigg[\big(\omega _{+,+-}-\sqrt{2}\ell ^{-1}\big)\mathbf{e}^{+}-\omega _{+,+-}\mathbf{e}^{-}-\omega _{+,+\bar{1}}\mathbf{e}^{\bar{1}}
\bigg]\wedge \mathbf{e}^{1}\ .
\end{eqnarray}

To proceed, note that ({\ref{extalg}}) implies that 
\begin{equation}
(\mathbf{e}^{+}+\mathbf{e}^{-})\wedge \mathrm{d}(\mathbf{e}^{+}+\mathbf{e}
^{-})=0,\qquad (\mathbf{e}^{+}-\mathbf{e}^{-})\wedge \mathrm{d}(\mathbf{e}^{+}-\mathbf{e}^{-})=0\ .
\end{equation}
Hence, there exist real functions $H,B,z,t$ such that 
\begin{equation}
\mathbf{e}^{+}={\frac{1}{\sqrt{2}}}\big(Hdz-Bdt),\qquad \mathbf{e}^{-}={\frac{1}{\sqrt{2}}}(Hdz+Bdt)\ .
\end{equation}
Next, note that ({\ref{gaugefs1}}) and ({\ref{extalg}}) imply that 
\begin{equation}
F={\frac{1}{\sqrt{2}}}d(\mathbf{e}^{+}+\mathbf{e}^{-})\ .
\end{equation}

On comparing this expression with ({\ref{gaugepot1}}), one finds that there
exists a real function $C$ such that 
\begin{eqnarray}
{\frac{1}{\sqrt{2}}}(\mathbf{e}^{+}+\mathbf{e}^{-}) &=&{\frac{\ell }{2}}\big(
\omega _{+,+-}\mathbf{e}^{+}-\omega _{-,+-}\mathbf{e}^{-}-\omega _{+,+1}
\mathbf{e}^{1}-\omega _{+,+\bar{1}}\mathbf{e}^{\bar{1}}\big)  \notag \\
&-&{\frac{\ell }{2}}d\log C\ .
\end{eqnarray}
Substituting this expression back into ({\ref{extalg}}) one finds 
\begin{equation}
d\mathbf{e}^{1}=d\log C\wedge \mathbf{e}^{1}
\end{equation}
and so there exist real functions $C$, $x$, $y$ such that 
\begin{equation}
\mathbf{e}^{1}={\frac{1}{\sqrt{2}}}C(dx+idy)\ .
\end{equation}

It is then straightforward to show that ({\ref{extalg}}) implies that 
\begin{equation}
H=C^{-1}f_{1}(z),\qquad B=Cf_{2}(t)
\end{equation}
where $f_{1}$ and $f_{2}$ are arbitrary functions of $z$, $t$.

By making appropriate $z$, $t$ co-ordinate transformations, one can without
loss of generality take $f_{1}=f_{2}=1$. Furthermore, ({\ref{extalg}})
implies that 
\begin{equation}
{\frac{\partial C}{\partial z}}=-\ell ^{-1}
\end{equation}
so that 
\begin{equation}
C=V-\ell ^{-1}z
\end{equation}
for $V=V(t,x,y)$. The metric and gauge field strength are then given by 
\begin{equation}
ds^{2}=(V-\ell ^{-1}z)^{2}(-dt^{2}+dx^{2}+dy^{2})+{\frac{1}{(V-\ell
^{-1}z)^{2}}}dz^{2}  \label{cosmp}
\end{equation}
and 
\begin{equation}
F=d\bigg({\frac{1}{(V-\ell ^{-1}z)}}dz\bigg)\ .
\end{equation}

Finally, we impose the gauge field equations $d\star F=0$, which imply that $V$ is harmonic on ${\mathbb{R}}^{1,2}$: 
\begin{equation}
\left( -{\frac{\partial ^{2}}{\partial t^{2}}}+{\frac{\partial ^{2}}{\partial x^{2}}}+{\frac{\partial ^{2}}{\partial y^{2}}}\right) V=0\ ,
\end{equation}
and we remark that, from the reasoning used in the previous sub-section,
this condition is sufficient to ensure that the Einstein equations hold
automatically. In the limit of $\ell ^{-1}\rightarrow 0,$ one recovers the
solutions found in \cite{fake}. Note that our cosmological solutions are
obtained by shifting the harmonic function independent of $z$ by a term
linear in $z.$ This is similar in the case of de Sitter solutions in which
the solution is obtained by shifting the time-independent harmonic function
by linear term in time \cite{ds, shift}. Cosmological de sitter extensions
of Majumdar-Papapetrou solutions were first considered in \cite{kt}.

\section{Conclusions}

Using spinorial geometry techniques, all pseudo-supersymmetric solutions of
minimal Anti de Sitter $N=2$, $D=4$ supergravity have been classified. There
are four classes of solutions:

\begin{itemize}
\item[(i)] The first class of solution has metric and field strength 
\begin{equation}
ds^{2}=2du\bigg[dv+\big(-\ell ^{-2}v^{2}+\Psi \big)du\bigg]+{\frac{2}{(1-2\ell ^{-2}z{\bar{z}})^{2}}}dzd{\bar{z}}  \label{grav}
\end{equation}
with 
\begin{equation}
F=-\ell ^{-1}dv\wedge du
\end{equation}
where $\Psi =\Psi (u,z,{\bar{z}})$ satisfies 
\begin{equation}
{\frac{\partial ^{2}\Psi }{\partial z\partial {\bar{z}}}}=0\ .
\end{equation}

This metric corresponds to gravitational waves propagating on $AdS_2 \times
H^2$, \cite{meessen, nariai, podol}.

\item[(ii)] The second class of solution is a type III Kundt solution, corresponding
to one of the geometries in the classification of \cite{Griffiths}. The metric is 
\begin{eqnarray}
ds^{2} &=&2du\bigg(dv+\big({\mathcal{F}}+v(\Psi -\sqrt{2}\ell \partial
_{x}\Psi )\big)du+\sqrt{2}\big(\ell ^{-1}v+\ell \Psi \big)dx\bigg)+ds_{2}^{2}
\notag \\
&&
\end{eqnarray}
where $\Psi $, ${\mathcal{F}}$ are functions of $u,x,y$; and 
\begin{equation}
ds_{2}^{2}={\frac{1}{2}}dx^{2}+{\frac{1}{2}}e^{-\sqrt{2}\ell ^{-1}x}dy^{2}
\end{equation}
is the metric on $H^{2}$, and 
\begin{equation}
F={\frac{\ell }{2}}d\Psi \wedge du\ .
\end{equation}
The function $\Psi $ is harmonic on $H^{2}$, and ${\mathcal{F}}$ satisfies 
\begin{eqnarray}
{\partial^2 {\cal{F}} \over \partial x^2}
+  e^{\sqrt{2} \ell^{-1}x} {\partial^2 {\cal{F}} \over \partial y^2}
+{1 \over \sqrt{2}} \ell^{-1} {\partial {\cal{F}} \over \partial x}
- \ell^{-2} {\cal{F}}
&=& -{\partial \Psi \over \partial u}
+4 \sqrt{2} \ell \Psi {\partial \Psi \over \partial x} - \Psi^2
\nonumber \\
&-&{3 \over 2} \ell^2 \bigg({\partial \Psi \over \partial x} \bigg)^2
+\ell \sqrt{2} {\partial^2 \Psi \over \partial x \partial u}
\nonumber \\
&+&{5 \over 2} \ell^2 e^{\sqrt{2} \ell^{-1}x}\bigg({\partial \Psi \over \partial y} \bigg)^2
-4 \ell^2 \Psi {\partial^2 \Psi \over \partial x^2} \ .
\nonumber \\
\end{eqnarray}

\item[(iii)] The third class of solution has metric 
\begin{equation}
ds^{2}=\bigg({\frac{\ell }{\sin \theta }}d\theta -\ell \cos \theta \ {%
\mathcal{B}}+\sqrt{2}\sin \theta \ \psi \bigg)^{2}+{\frac{1}{\sin ^{2}\theta 
}}ds_{LGT}^{2}
\end{equation}
where $ds_{LGT}^{2}$ is a $\theta $-independent metric on a 3-dimensional
Lorentzian manifold which has a Lorentzian Gauduchon-Tod structure. The
3-manifold $LGT$ admits a $\theta $-independent basis $\mathbf{E}^{i}$ and a 
$\theta $-independent 1-form ${\mathcal{B}}$ satisfying 
\begin{equation}
d\mathbf{E}^{i}={\mathcal{B}}\wedge \mathbf{E}^{i}+2\ell ^{-1}\star _{3}\mathbf{E}^{i}
\end{equation}
together with a $\theta $-independent 1-form $\psi $ satisfying 
\begin{equation}
d\psi +{\mathcal{B}}\wedge \psi +2\ell ^{-1}\star _{3}\psi =0\ .
\end{equation}
The gauge field strength is 
\begin{equation}
F=d\bigg(-{\frac{\ell }{2}}\cos 2\theta \ {\mathcal{B}}+{\frac{1}{\sqrt{2}}}
\sin 2\theta \ \psi \bigg)\ .
\end{equation}

\item[(iv)] The fourth class of solution has metric 
\begin{equation}
ds^{2}=(V-\ell ^{-1}z)^{2}ds^{2}(\mathbb{R}^{2,1})+{\frac{1}{(V-\ell
^{-1}z)^{2}}}dz^{2}
\end{equation}
and 
\begin{equation}
F=d\bigg({\frac{1}{(V-\ell ^{-1}z)}}dz\bigg)\ .
\end{equation}
The function $V$ does not depend on $z$, and satisfies the wave equation on ${\mathbb{R}}^{2,1}$.
\end{itemize}

\appendix

\section{The Linear System}

In this appendix we present the decomposition of the Killing spinor equation
acting on the spinor $\epsilon =\lambda 1+\mu ^{i}e_{i}$; we obtain the
following conditions:

{\ 
\begin{eqnarray}
\partial _{+}\lambda +\lambda \left( -{\frac{1}{2}}\omega _{+,+-}-{\frac{1}{2}}\omega _{+,1{\bar{1}}}-\ell ^{-1}A_{+}\right) -\frac{1}{\sqrt{2}}\mu
^{2}\left( F_{+-}+F_{1{\bar{1}}}-\ell ^{-1}\right) &=&0,  \notag \\
\partial _{+}\mu ^{1}+\mu ^{1}\left( -{\frac{1}{2}}\omega _{+,+-}+{\frac{1}{2}}\omega _{+,1{\bar{1}}}-\ell ^{-1}A_{+}\right) -\omega _{+,-1}\mu ^{2} &=&0,
\notag \\
\partial _{+}\mu ^{2}+\omega _{+,+{\bar{1}}}\mu ^{1}+\mu ^{2}\left( {\frac{1}{2}}\omega _{+,+-}-{\frac{1}{2}}\omega _{+,1{\bar{1}}}-\ell
^{-1}A_{+}\right) &=&0,  \notag \\
\omega _{+,+1}\lambda +\sqrt{2}F_{+1}\mu ^{2} &=&0,  \notag \\
\partial _{-}\lambda +\lambda \left( -{\frac{1}{2}}\omega _{-,+-}-{\frac{1}{2}}\omega _{-,1{\bar{1}}}-\ell ^{-1}A_{-}\right) -\sqrt{2}F_{-{\bar{1}}}\mu
^{1} &=&0,  \notag \\
\partial _{-}\mu ^{1}-\sqrt{2}F_{-1}\lambda +\mu ^{1}\left( -{\frac{1}{2}}\omega _{-,+-}+{\frac{1}{2}}\omega _{-,1{\bar{1}}}-\ell ^{-1}A_{-}\right)
-\omega _{-,-1}\mu ^{2} &=&0,  \notag \\
\partial _{-}\mu ^{2}+\frac{\lambda }{\sqrt{2}}\left( F_{+-}-F_{1{\bar{1}}}+\ell ^{-1}\right) +\omega _{-,+{\bar{1}}}\mu ^{1}+\mu ^{2}\left( {\frac{1}{2}}\omega _{-,+-}-{\frac{1}{2}}\omega _{-,1{\bar{1}}}-\ell ^{-1}A_{-}\right)
&=&0,  \notag \\
-\omega _{-,+1}\lambda -\frac{1}{\sqrt{2}}\mu ^{1}\left( F_{+-}+F_{1{\bar{1}}}+\ell ^{-1}\right) &=&0,  \notag \\
\partial _{1}\lambda +\lambda \left( -{\frac{1}{2}}\omega _{1,+-}-{\frac{1}{2}}\omega _{1,1{\bar{1}}}-\ell ^{-1}A_{1}\right) -{\frac{1}{\sqrt{2}}}\mu
^{1}\left( F_{+-}+F_{1{\bar{1}}}-\ell ^{-1}\right) &=&0,  \notag \\
\partial _{1}\mu ^{1}+\mu ^{1}\left( -{\frac{1}{2}}\omega _{1,+-}+{\frac{1}{2}}\omega _{1,1{\bar{1}}}-\ell ^{-1}A_{1}\right) -\omega _{1,-1}\mu ^{2} &=&0,
\notag \\
\partial _{1}\mu ^{2}+\omega _{1,+{\bar{1}}}\mu ^{1}+\mu ^{2}\left( {\frac{1}{2}}\omega _{1,+-}-{\frac{1}{2}}\omega _{1,1{\bar{1}}}-\ell
^{-1}A_{1}\right) &=&0,  \notag \\
\omega _{1,+1}\lambda +\sqrt{2}F_{+1}\mu ^{1} &=&0,  \notag \\
\partial _{\bar{1}}\lambda +\lambda \left( -{\frac{1}{2}}\omega _{{\bar{1}},+-}-{\frac{1}{2}}\omega _{{\bar{1}},1{\bar{1}}}-\ell ^{-1}A_{{\bar{1}}}\right) +\sqrt{2}F_{-{\bar{1}}}\mu ^{2} &=&0,  \notag \\
\partial _{\bar{1}}\mu ^{1}+\frac{\lambda }{\sqrt{2}}\left( -F_{+-}+F_{1{\bar{1}}}+\ell ^{-1}\right) +\mu ^{1}\left( -{\frac{1}{2}}\omega _{{\bar{1}}
,+-}+{\frac{1}{2}}\omega _{{\bar{1}},1{\bar{1}}}-\ell ^{-1}A_{\bar{1}}\right) -\omega _{{\bar{1}},-1}\mu ^{2} &=&0,  \notag \\
\partial _{\bar{1}}\mu ^{2}+\sqrt{2}F_{+{\bar{1}}}\lambda +\omega _{{\bar{1}}
,+{\bar{1}}}\mu ^{1}+\mu ^{2}\left( {\frac{1}{2}}\omega _{{\bar{1}},+-}-{\frac{1}{2}}\omega _{{\bar{1}},1{\bar{1}}}-\ell ^{-1}A_{\bar{1}}\right) &=&0,
\notag \\
-\omega _{{\bar{1}},+1}\lambda +\frac{1}{\sqrt{2}}\mu ^{2}\left( F_{+-}+F_{1{\bar{1}}}+\ell ^{-1}\right) &=&0\ .
\nonumber \\
\end{eqnarray}
}

{\flushleft{\textbf{Acknowledgements:}}} JG is supported by the STFC
Consolidated Grant ST/L000490/1. JG thanks the American University of
Beirut for hospitality when part of this work was undertaken. 
The work of WS is supported in part by
the National Science Foundation under grant number PHY-1620505.
The authors would like to thank M. Ortaggio for useful discussions
concerning the relationship between the solution constructed in Section 4.2, and
the classification in \cite{Griffiths}.

\end{document}